\documentclass[showpacs,preprintnumbers,amsmath,amssymb,aps,twocolumn]{revtex4}% uncomment for publication layout (endfloats for the floats to the end)

\usepackage{bm}%
\usepackage{graphicx}
\usepackage{color}
\usepackage{amsmath}
\usepackage{xspace}

\newcommand{\InP}{InP:Fe\xspace}
\newcommand{\LN}{LiNbO$_3$\xspace}
\newcommand{\CdZnTe}{CdZnTe\xspace}
\newcommand{\SBN}{Sr$_x$Ba$_{1-x}$Nb$_2$O$_6$\xspace}
\newcommand{\BTO}{Bi$_{12}$TiO$_{20}$\xspace}

\newcommand{\AffilMops}{\affiliation{Laboratoire Mat\'eriaux Optiques, Photonique et Syst\`emes \\ Unit\'e de Recherche Commune \`a l'Universit\'e Paul Verlaine -- Metz et Sup\'elec - CNRS UMR 7132 \\ 2, rue Edouard Belin, 57070 Metz, France}}
\newcommand{\AffilLopmd}{\affiliation{Institut FEMTO-ST Universit\'e de Franche Comt\'e D\'epartement d'optique - CNRS UMR 6174\\ UFR Sciences et Techniques Route de Gray 25030 Besan\c{c}on, France }}
\newcommand{\AffilPoma}{\affiliation{Laboratoire des Propri\'et\'es Optiques des Mat\'eriaux et Applications - CNRS UMR 6136\\ UFR Sciences, 2 Bd Lavoisier 49045 Angers, France}}

\newcommand{\Micron}{\mu\mathrm{m}}
\newcommand{\Milli}{\mathrm{m}}
\newcommand{\cm}{\mathrm{cm}}
\newcommand{\Celsius}{^{\circ}\mathrm{C}}
\newcommand{\Watt}{\mathrm{W}}
\newcommand{\kV}{\mathrm{kV}}
\newcommand{\mm}{\mathrm{mm}}

\newcommand{\EtAl}{\emph{et. al.}\xspace}
\newcommand{\NIR}{$1.06\Micron$\xspace}
\newcommand{\FIR}{$1.56\Micron$\xspace}
\newcommand{\modelone}{model I\xspace}
\newcommand{\modeltwo}{model II\xspace}
\newcommand{\modelthree}{model III\xspace}
\newcommand{\Bound}[1]{#1_0}
\newcommand{\Frac}[2]{{#1}/{#2}}
\newcommand{\Dark}[2]{\Frac{e^{th}_{#1} {#1}_{\Bound T}}{c_{#1} {#2}_{\Bound T}}}
\newcommand{\Current}[1]{e\mu_{#1} \Bound{#1} \Bound{E}}
\newcommand{\Ires}{I_r}
\newcommand{\Idark}{I_d}
\newcommand{\Fe}[1]{\mathrm{\left[ Fe^{#1+}\right]}}
\newcommand{\SubSection}[1]{}

\newcommand{\Hfig}{20mm}
\newcommand{\ToThe}{\times}

\newcommand{\TextColor}[2]{{#2}}

\begin{document}

\title {Roles of resonance and dark irradiance for infrared photorefractive self-focusing and solitons in bi-polar \InP}
\author{N.Fressengeas}
\AffilMops
\author{N.Khelfaoui}
\AffilMops
\author{C.Dan}
\AffilMops
\author{D.Wolfersberger}
\AffilMops
\author{G.Montemezzani}
\AffilMops
\author{H.Leblond}
\AffilPoma
\author{M.Chauvet}
\AffilLopmd

\pacs{42.70.Nq, 42.65.Tg}
\begin{abstract}
%Semi-conductors such as \InP are attractive materials for photorefractive solitons, because of their sensitivity to infrared telecommunication wavelengths and their high carrier mobility offering a high speed photorefractive effect.
This paper shows experimental evidence of photorefractive steady state self-focusing in \InP for a wide range of intensities, at both $1.06$ and $1.55\Micron$. To explain those results, it is  shown that despite the bi-polar nature of \InP where one photocarrier and one thermal carrier are to be considered, the long standing one photocarrier model for photorefractive solitons can be usefully applied. The relationship between the dark irradiance stemming out of this model and the known resonance intensity is then discussed.
\end{abstract}

\maketitle

\section{Introduction}
Photorefractive self-focusing and spatial solitons have been up to now the object of more than a decade of growing interest, both experimentally and theoretically~\cite{seg92prl923,dur93prl533,cro93josab446,itu94apl408}. During this decade, solitary waves induced by the photorefractive effect have been shown to interact with each other showing specific behaviors upon collision~\cite{men97ol448} like birth and fusion~\cite{kro97ol369}. When a spatial solitary wave propagates in a photorefractive medium, it induces an optical waveguide that does not disappear instantly when the beam is turned off~\cite{mor95ol2066,shi96ol931}, leading to optically induced optical components such as  junctions~\cite{che96ol716}, couplers~\cite{lan99ol475} and even optically induced soliton arrays and photonic lattices~\cite{des06oe2851,pet03ol438,fle05oe1780}.

The attraction towards photorefractive solitons and the waveguides they induce owes a lot to their potentialities in optically inducing photonic circuitry in the bulk of a material, on the one hand, and to their reconfiguration properties, on the other hand. The former relies on the material property of retaining the waveguide once the beam is shut off, like, for instance, \LN~\cite{cha04apl2193}. The latter, on the contrary, relies on the possibility to rapidly screen out an applied field with a laser beam, to give rise to a waveguide, and to erase this waveguide as rapidly. Up to now and to our knowledge, the characteristic build-up and erase times are on the order of the second in the widely used \SBN~\cite{wes01pre36613} or sillenite \BTO~\cite{fre99jap2062}, or up to the minute in \LN~\cite{cha01ol1344}, all of these experiments having been conducted at visible wavelengths to which most of the cited materials are sensitive. Shorter response times around the millisecond have been obtained in semiconductor \CdZnTe~\cite{sch02ol1229} and around the nanosecond  for visible ultra-high intensities in \BTO~\cite{wol00pre}.

If one wants to target optical networking as a potential application of the dynamically reconfigurable photorefractive soliton technology, build-up and reconfiguration times below the millisecond is to be achieved ideally at infrared wavelengths compatible with optical fiber networks. Thanks to their high carrier mobility and despite their low electro-optic coefficient, photorefractive semi-conductors such as \InP  might respond to this demand.
Charge transport, in this material, has a bi-polar nature with photoexcited holes co-existing with thermally excited electrons. This yields a peculiar resonance of the photorefractive Two Wave Mixing (TWM) gain as a function of intensity\cite{Idr87oc317,mai88ol657,pic89ol1362,Pic89ap3798}.
%Their ability to sustain photorefractive spatial solitons has been demonstrated experimentally a decade ago~\cite{cha97apl2499,Cha96ol1333}. However, the photorefractive physics at work to generate them are more complicated than in the previously cited experiments. Indeed, \InP semiconductor bipolar nature compels us to consider both charge carriers, thermally-excited electrons and photo-generated holes~\cite{Idr87oc317,mai88ol657,pic89ol1362,Pic89ap3798}. Since then, one attempt has been carried out at theoretically describing the obtained experimental results~\cite{uzd01ol1547}.

%Thanks to this previous work, we intended to pursue the steady-state investigations in \InP. We thus propose in this paper a new series of  experiments and a novel model of photorefractive charge transport in \InP for application to self-focusing at steady state,  in the light of which we will reconsider the conclusions of the above cited experiments and modelling, particularly concerning the predicted and observed resonance intensity.

Owing to its bipolar transport character, the physics behind spatial soliton propagation in \InP is more complicated than in other materials. It has already been the subject of theoretical analysis\cite{uzd01ol1547}. In this work, we present a series of recent investigations on photorefractive self-focusing in \InP at infrared wavelengths and we propose a model that takes into account the transport of both charge carriers. Unlike in previous work\cite{cha97apl2499,Cha96ol1333,uzd01ol1547}, the experimental results in our crystal samples and the corresponding theoretical prediction do not show any clear evidence of an influence of the intensity resonance of the bi-polar photorefractive effect, thus indicating that photorefractive self-focusing behaves to a large extent like in the case of the well-known single-carrier model\cite{chr95josab1628}. Hint is finally provided that the reason why qualitatively different behaviors are observed in different samples might be a varying oxydo-reduction state.

\section{Experimental evidence of the role played by dark irradiance}
\SubSection{Two Wave Mixing measurement of the resonance intensity}
The bi-polar photorefractive resonance mechanism was  first evidenced experimentally by Mainguet and Idrissi \EtAl~\cite{mai88ol657,Idr87oc317} and then defined by Picoli \EtAl~\cite{pic89ol1362,Pic89ap3798} as the mean intensity at which the Two Wave Mixing (TWM) gain is greatly enhanced. It corresponds to the intensity at which the intensity dependent hole generation rate is equal to the temperature dependent electron generation rate. The resonance intensity is thus expected to depend on the temperature and has been evidenced so. It has also been shown that this TWM enhancement corresponds to a $\pi/2$ phase shift between the index grating and the optical intensity grating, and to a corresponding enhancement of the index modulation amplitude.

As a preliminary characterization of our samples, we evidenced and measured the resonance intensity in the samples  used in our experiments. The iron content of the various samples we tested varied from $8\ToThe10^{16}\cm^{-3}$ to $2\ToThe10^{17}\cm^{-3}$, as measured by Secondary Ion Mass Spectroscopy. Similar behavior were observed in all samples. The results  presented in this paper were obtained with a $2\ToThe10^{17}\cm^{-3}$ Fe doped sample.

The resonance intensity has been measured with a standard TWM setup~\cite{Idr87oc317} as described in~\cite{khel06oc169}, for various temperatures, and particularly at $20\Celsius$, the temperature at which further self-focusing experimental results have been obtained. The measured resonance intensity at $20\Celsius$ is $300\Milli\Watt/\cm^2$ at a wavelength of  \NIR and $3\Watt/\cm^2$ at a wavelength of \FIR. This wavelength disparity is explained by the wavelength dependent photo-excitation cross sections.

\SubSection{Experimental setup}

The experiments were conducted independently at both \NIR and \FIR wavelengths, the latter because it corresponds to a  telecom wavelength and the former for \InP large sensitivity to it. The \NIR wavelength is delivered by a solid state diode pumped Nd:YAG laser whereas the \FIR beam is produced by a laser diode.

The  beam is focused down to a $30\Micron$ waist onto the entrance face of a temperature regulated \InP sample. The beam propagates along the  $<\overline{1}10>$ direction and  is linearly polarised along $<110>$ with an electric field of $10\kV/\cm$ applied along $<001>$. The crystal dimensions are $5\times5\times12\mm^3$ along the directions $<001><110> \mathrm{and} <\overline{1}10>$, respectively.

The steady state beam profile on the sample output face is observed through a microscope objective and a silicon CCD camera. The camera is sensitive enough for the \NIR wavelength but requires to be equipped with a luminescent phosphorus converter to achieve \FIR observations. The observed image is then gamma corrected to account for the conversion non-linearity.

\SubSection{Steady-state self focusing}

\begin{figure}
\begin{tabular}{ccc}
 \includegraphics[height=\Hfig,angle=180]{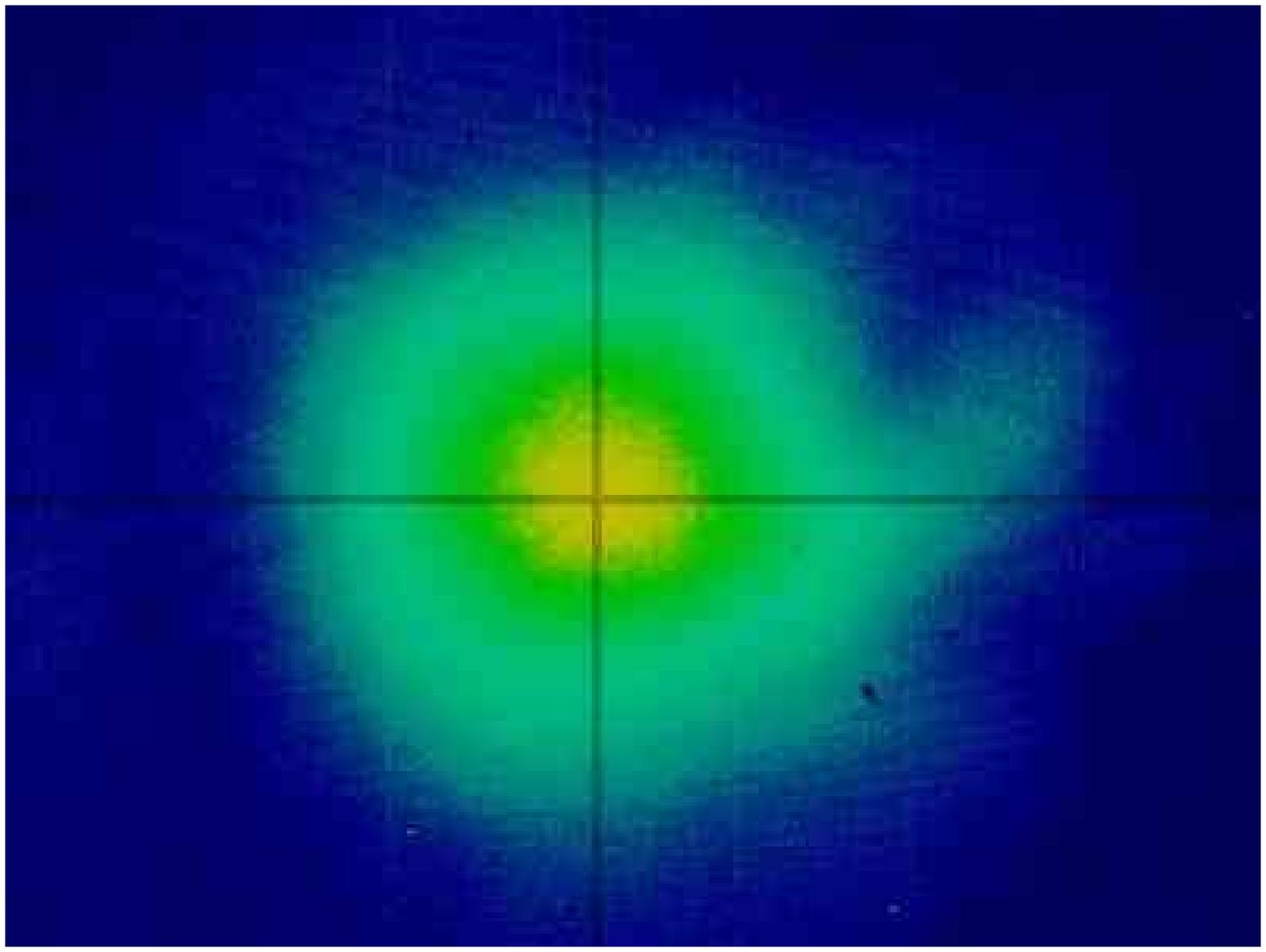}\hfill&
 \includegraphics[height=\Hfig,angle=180]{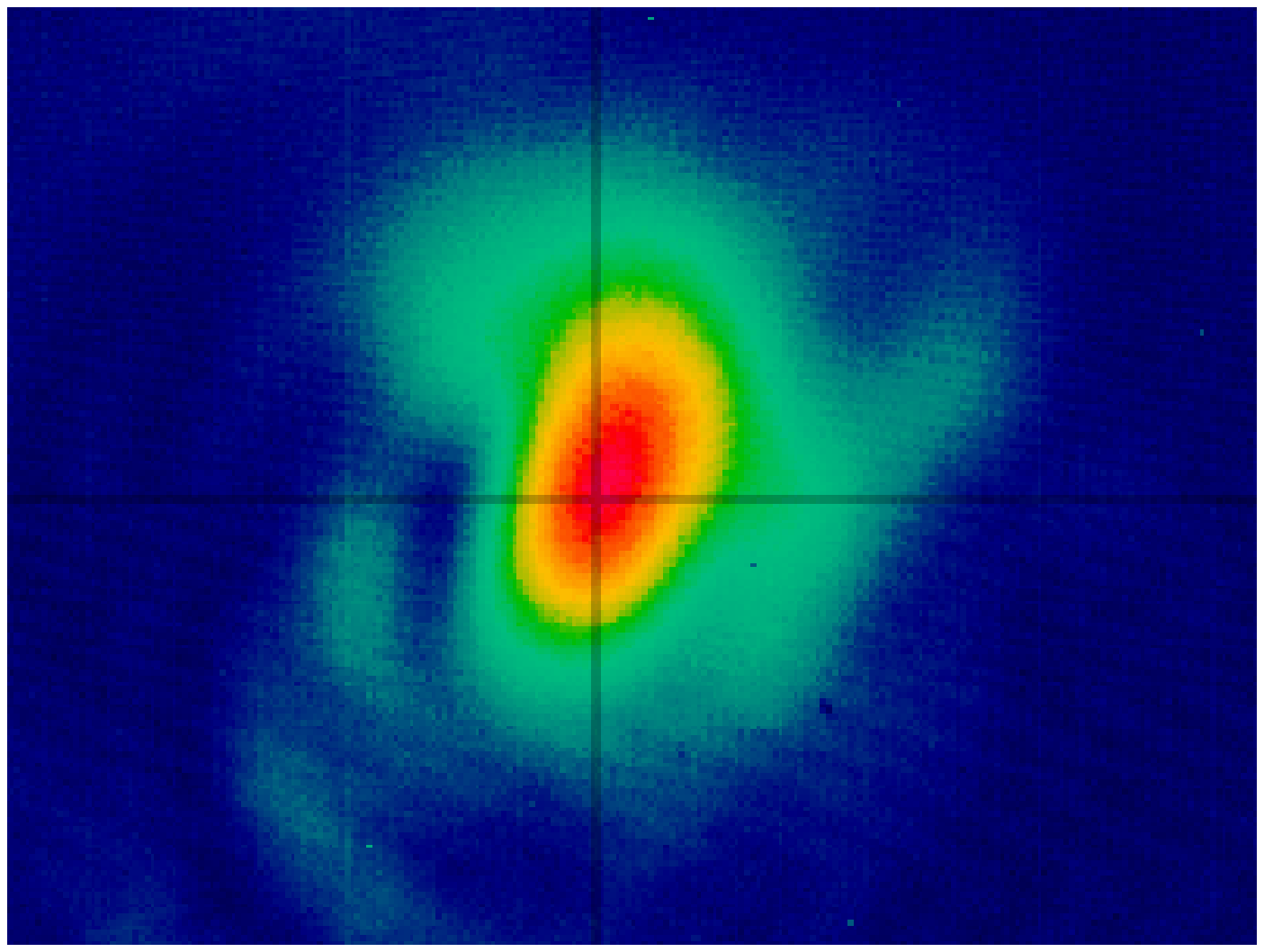}\hfill&
 \includegraphics[height=\Hfig,angle=180]{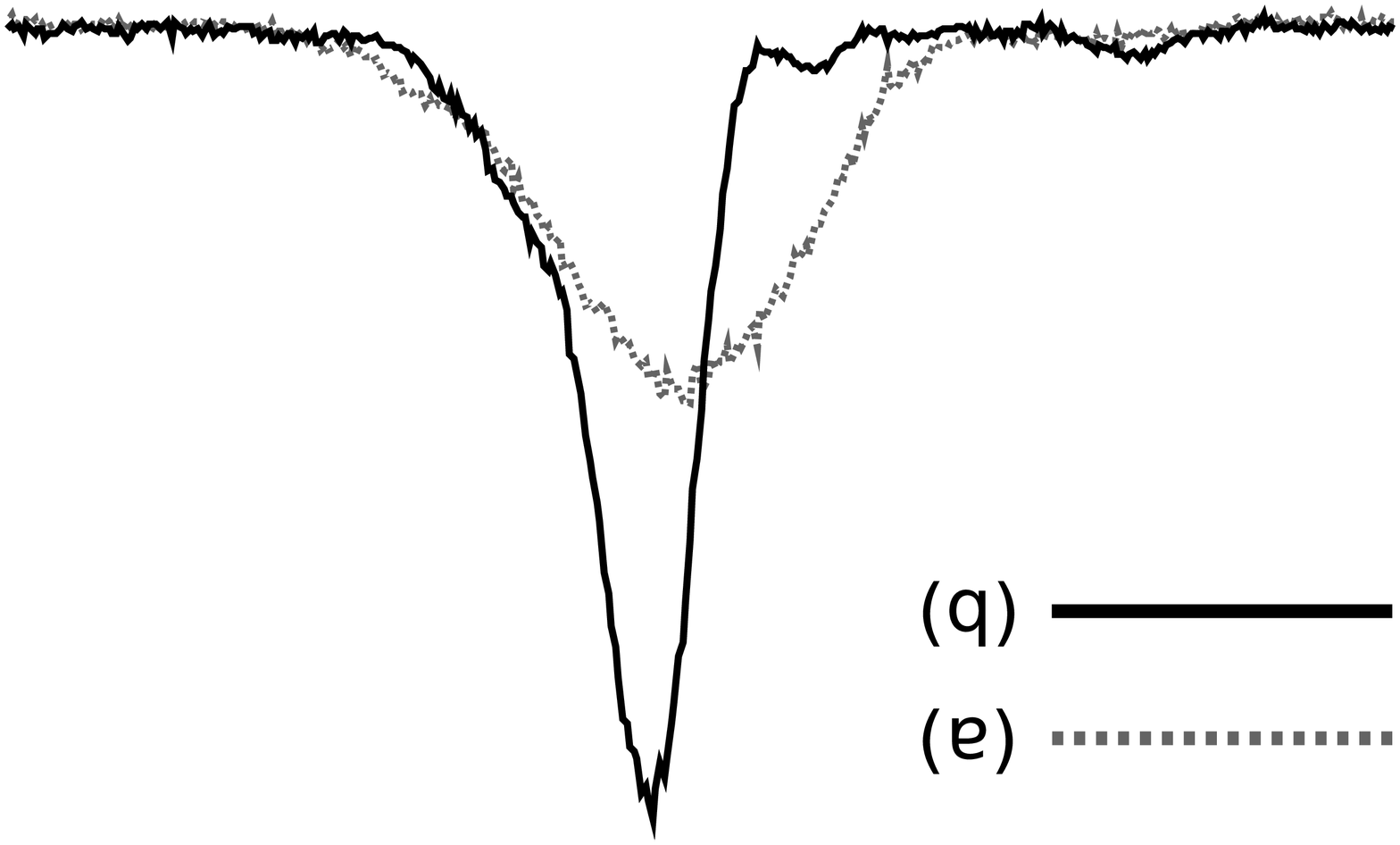}\\
(a)&(b)&(c)\\
 \includegraphics[height=\Hfig,angle=180]{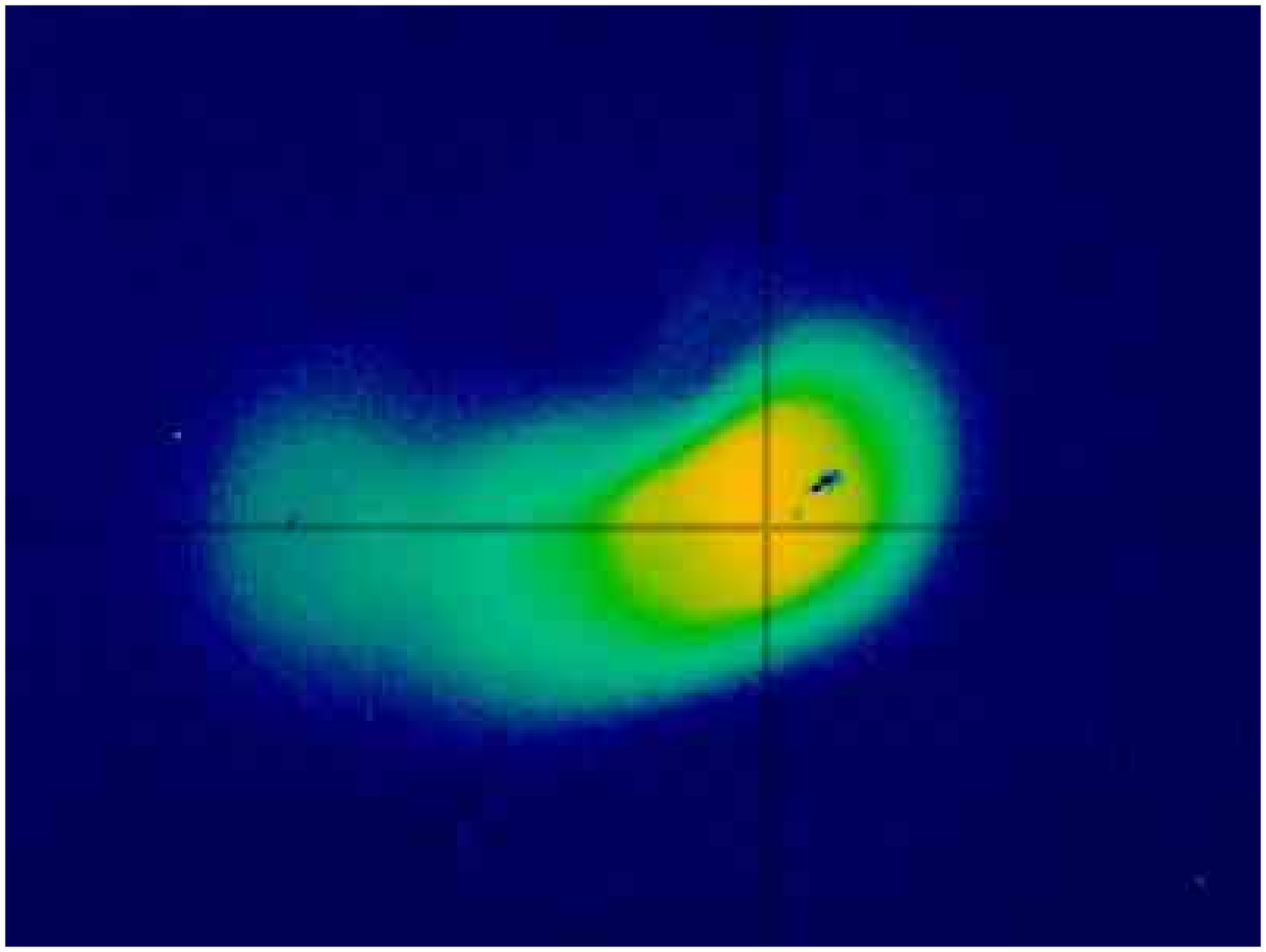}\hfill&
 \includegraphics[height=\Hfig,angle=180]{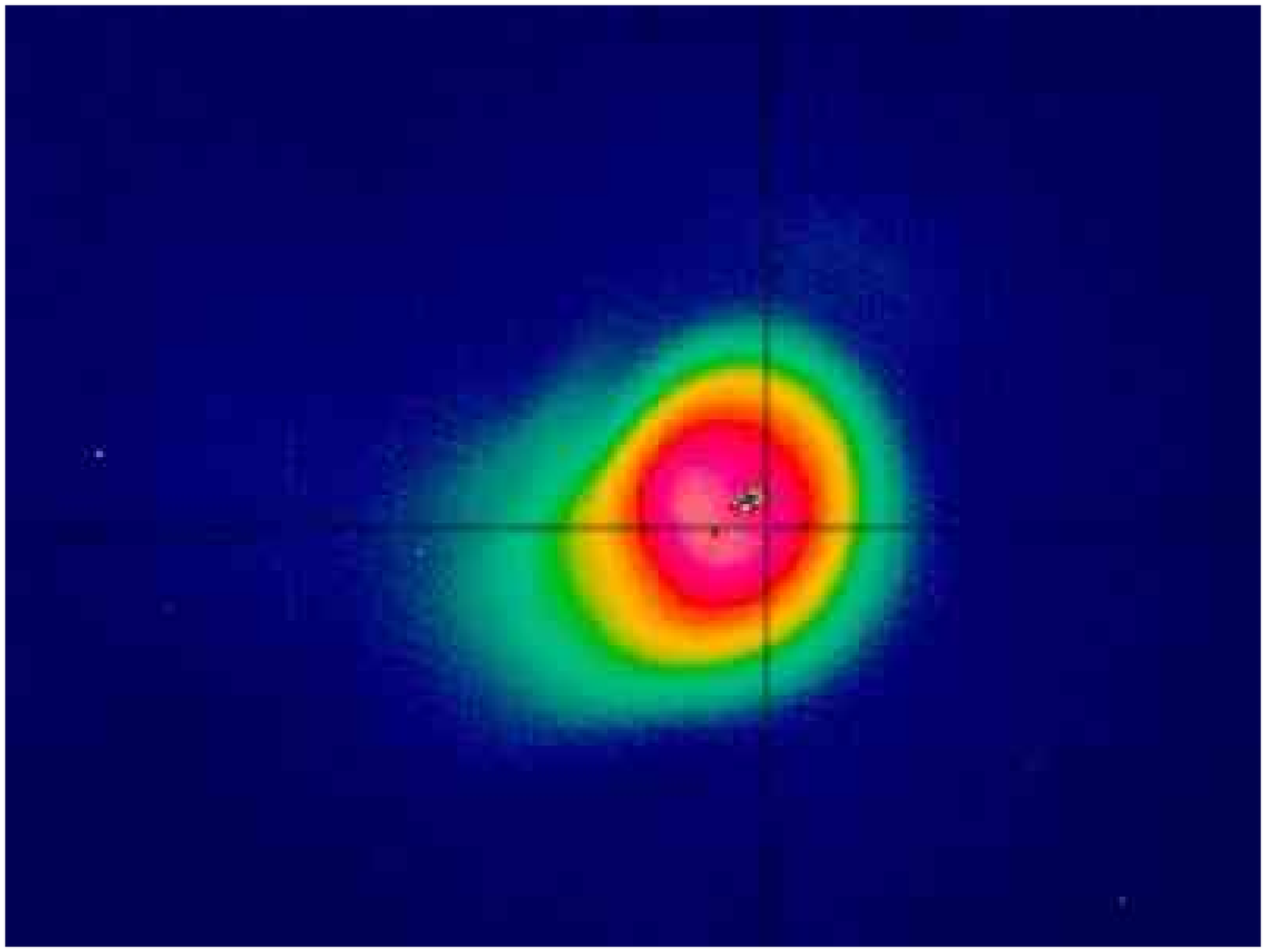}\hfill&
 \includegraphics[height=\Hfig,angle=180]{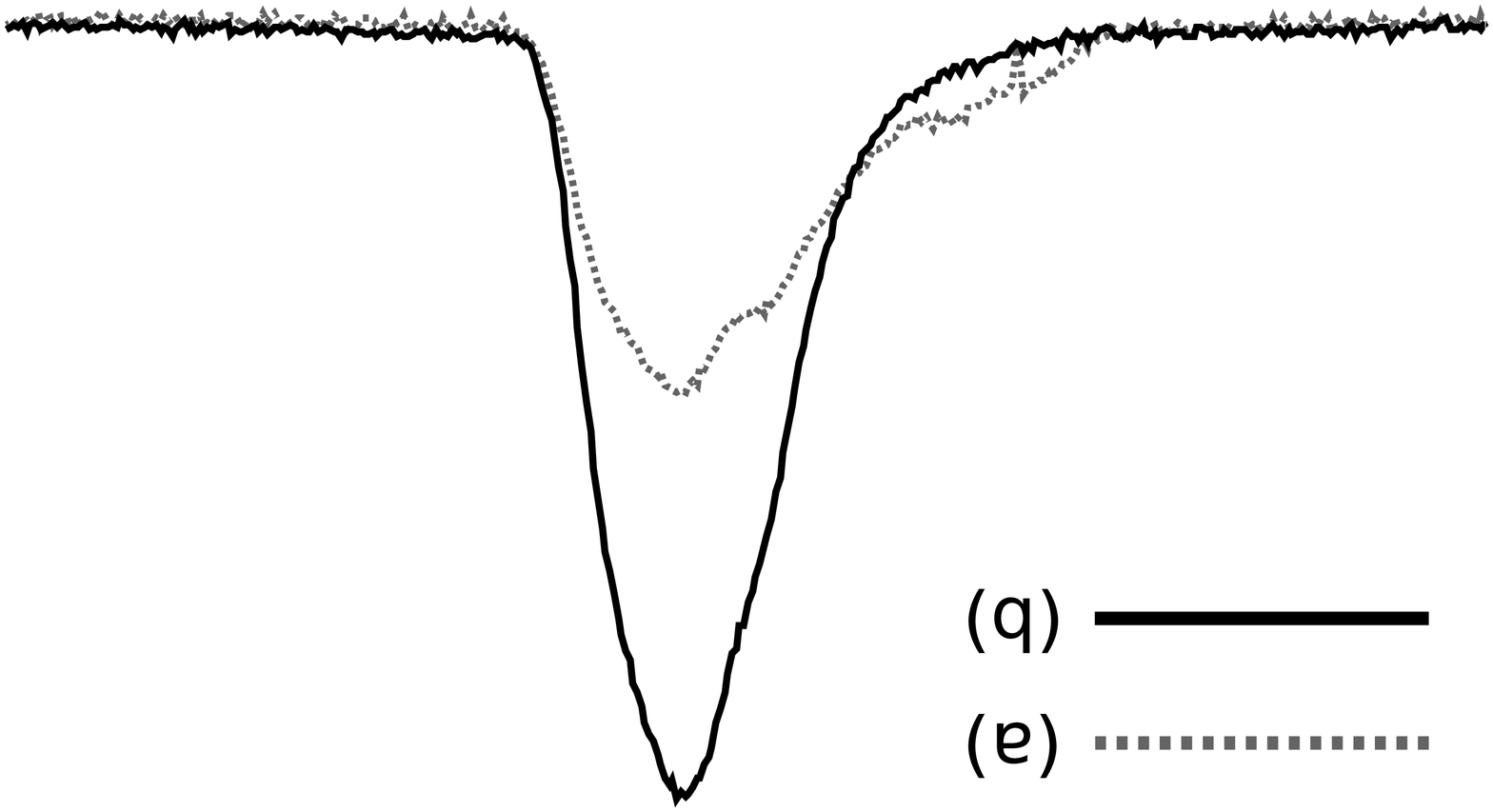}\\
(a)&(b)&(c)\\
\end{tabular}
\caption{(Color online) Steady state output beam profile for a peak intensity of
 $190\Milli\Watt/\cm^2$ at \NIR  (top) and a peak intensity of $8.7\Watt/\cm^2$ at \FIR (bottom), for no applied field (a) and a $10\kV/\cm$ horizontally applied field (b). The images total width is $200\Micron$. The non Gaussian shape of the \FIR beam in absence of the applied field is due to the multimode character of the laser itself. The curves (c) are the horizontal beam profiles (a.u.) at beam maximum.}\label{fig:focusimage}
\end{figure}

\begin{figure}
 \includegraphics[width=\linewidth]{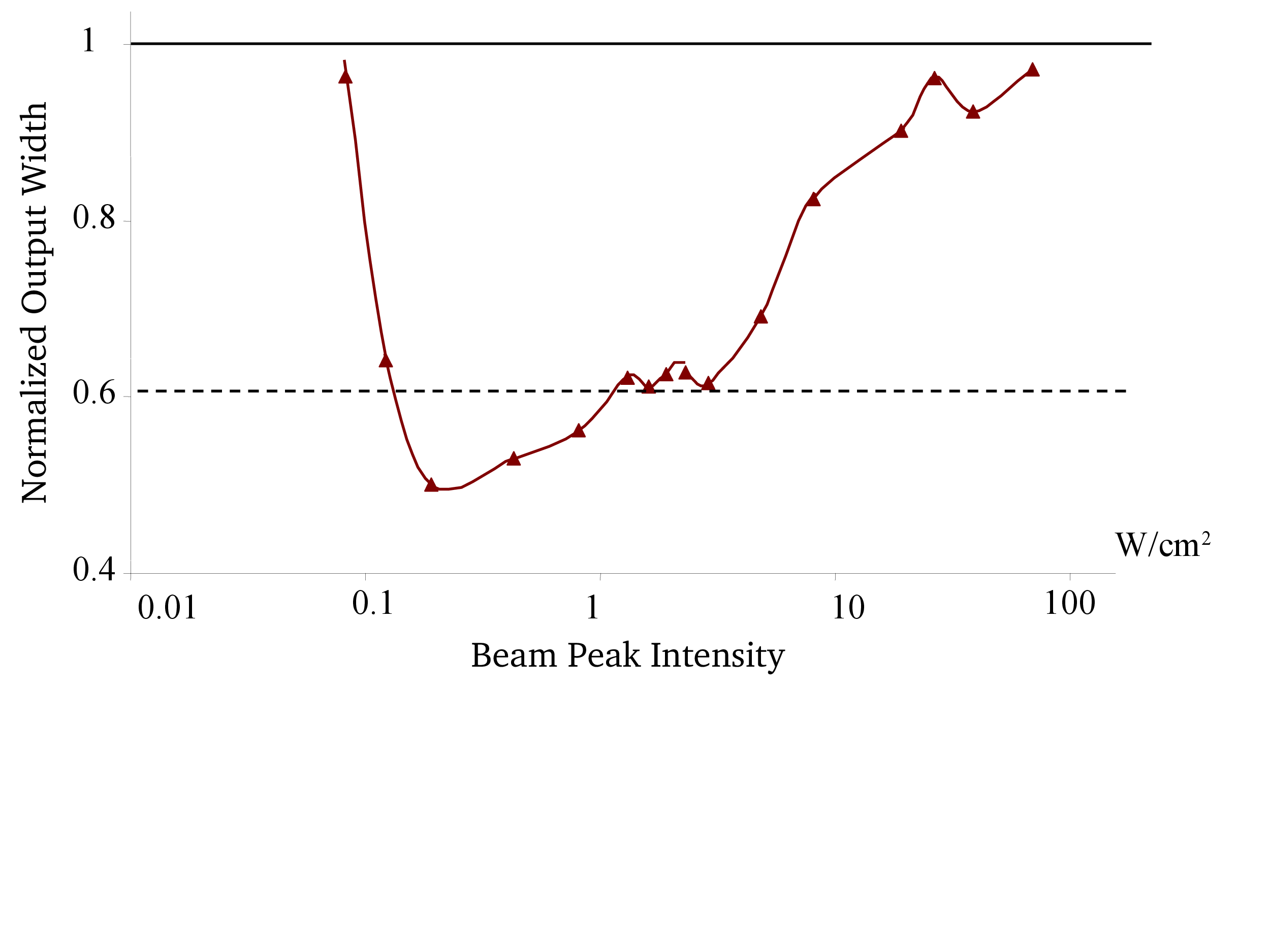}
 \includegraphics[width=\linewidth]{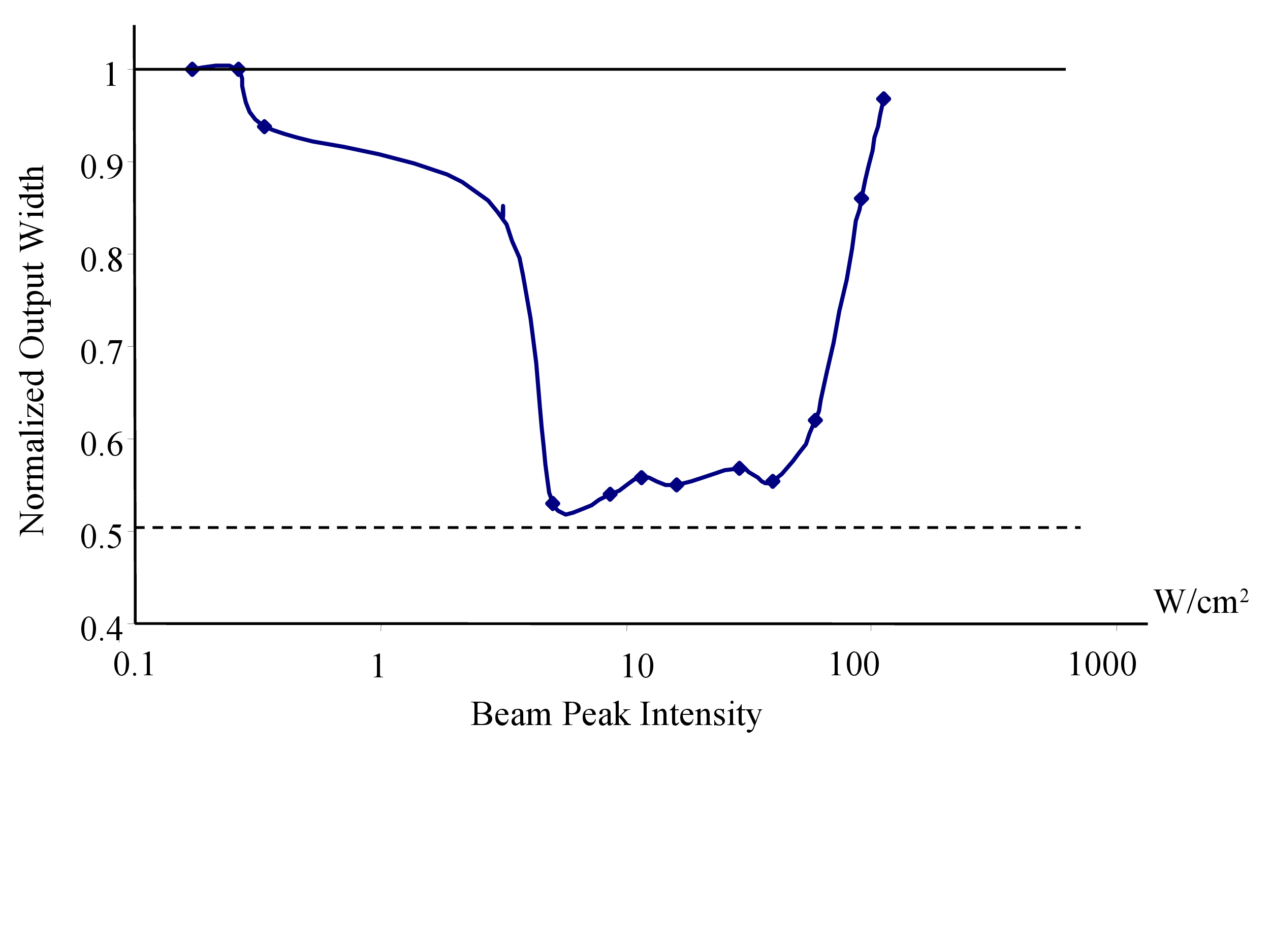}
\caption{Output beam width as a function of input beam peak intensity at \NIR (top) and \FIR (bottom) wavelengths. The beam diameter is normalized to the linearly diffracted output beam diameter. The horizontal dotted line represents an output beam as wide as the input one, suggesting a spatial soliton like propagation.}\label{fig:cloche}
\end{figure}

The output beam profile is observed at steady state for various intensities, at a temperature of $20\Celsius$. The best result giving the strongest focusing is shown in figure \ref{fig:focusimage}(b) for both \NIR and \FIR wavelengths.  Varying the input intensity reveals that the steady-state self-focusing phenomenon vanishes at low and high intensities and reaches  an optimum focusing for an intermediate intensity, when the output beam width reaches a minimum.

This is quantified on figure \ref{fig:cloche} where the normalized horizontal output beam  diameter is plotted as a function of the input beam peak intensity for an applied field of $10\kV/\cm$. Linear diffraction is observed at steady state for small and large intensities. A minimum output beam diameter is reached for an intermediate intensity of approximately $210\Milli\Watt/\cm^2$ at \NIR and $7.5\Watt/\cm^2$ at \FIR.
A noticeable characteristic of both curves in figure \ref{fig:cloche} is that the  upside down bell-shape is \emph{attracted} by the dotted line when it approaches it, \emph{i.e.} when the output beam diameter approaches the input one. This evidences a domain of stability of the beam width (from $1$ to $4\Watt/\cm^2$ at \NIR and from $5$ to $50\Watt/\cm^2$ at \FIR). The horizontal beam width associated to these domains roughly corresponds to the input beam width. This may suggest that the beam evolution is \emph{attracted} towards a solitary wave solution
% This can be interpreted as some kind of stability of a spatial solitary wave. Indeed, for a given input intensity, the conditions are reached for a spatial soliton to propagate. This suggests that spatial solitary wave does indeed propagate for a range of intensities: from $1$ to $4\Watt/\cm^2$ at \NIR and from $5$ to $50\Watt/\cm^2$ at \FIR.
%However, figure \ref{fig:cloche} shows that if the input intensity is close to the spatial-soliton condition, a spatial soliton can propagate for a given input intensity range. Let us note here that this observed stability suggests that a spatial soliton does indeed propagate in our \InP sample.

We note that the width stability intensity range is much wider at \FIR than at \NIR: this is due to the fact that in the former case, the intensity that fullfills soliton propagation conditions is much nearer to the minimum of the reverse bell-shaped curve.
Furthermore, one has to keep in mind that the beam intensity decreases with propagation, owing to absorption, --- $2\cm^{-1}$ at \NIR and  $0.5\cm^{-1}$ at \FIR--- so that the soliton propagation conditions cannot be fullfilled precisely all the way through the crystal, thus strengthening the need for stability if a spatial soliton is to propagate.

%\SubSection{Steady-state beam bending}

\begin{figure}
 \includegraphics[width=\linewidth]{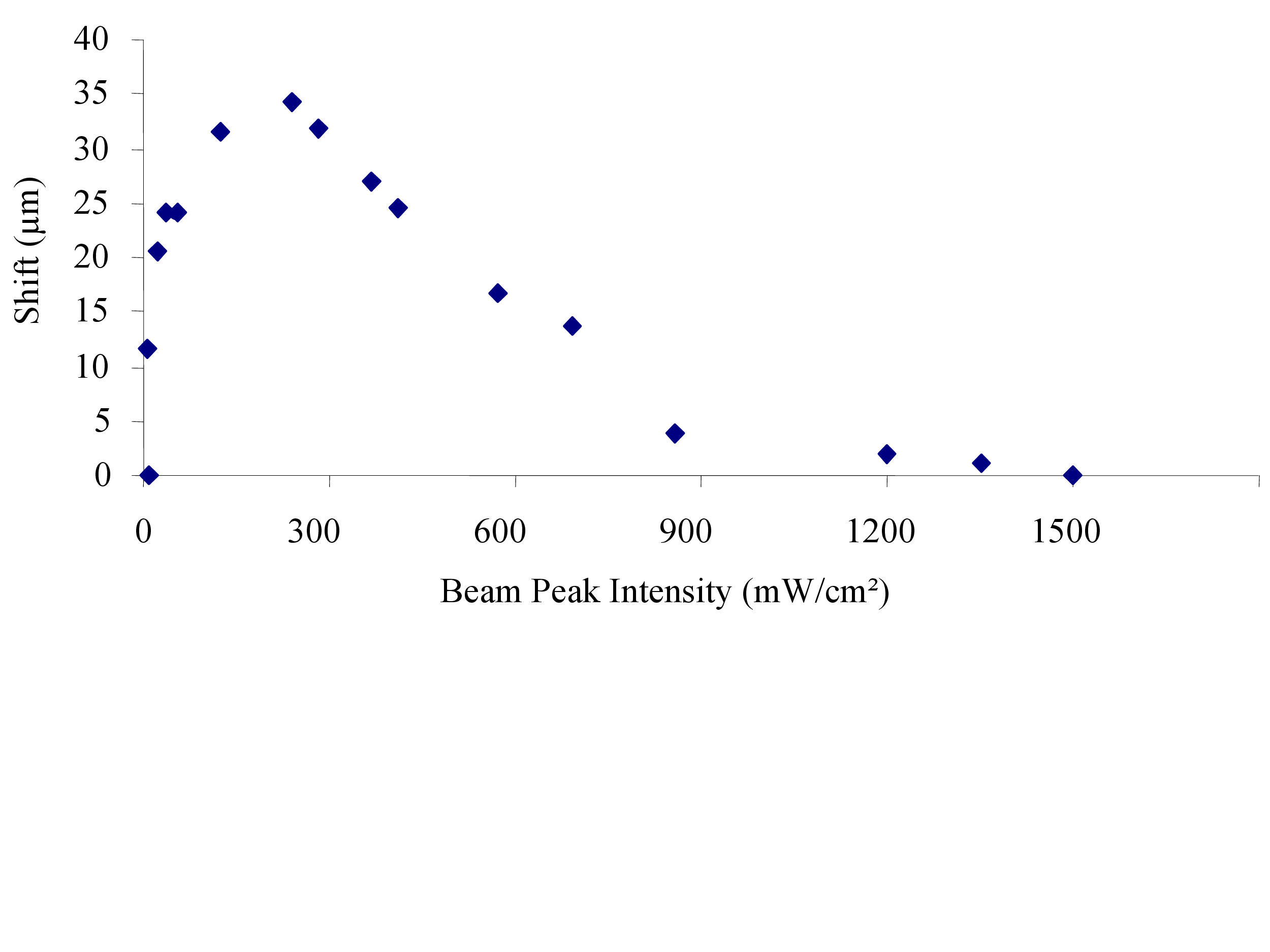}
 \includegraphics[width=\linewidth]{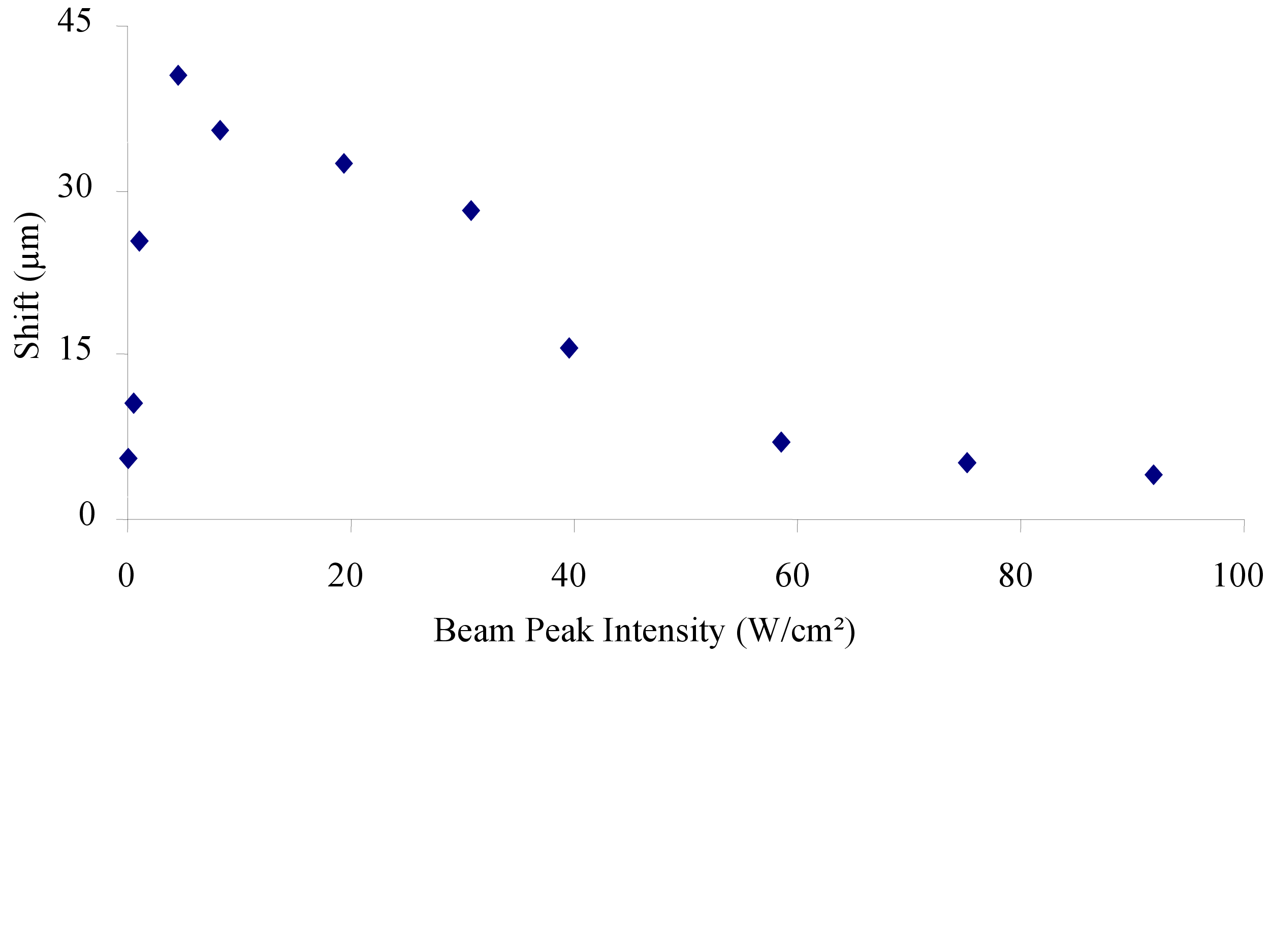}
\caption{Output beam horizontal shift as a function of input beam peak intensity at \NIR (top) and \FIR (bottom) wavelengths.}\label{fig:bending}
\end{figure}

As in many other materials, self-focusing in \InP comes together with beam bending, as is evidenced by figure \ref{fig:focusimage} and particularly \ref{fig:focusimage}(c). The steady-state beam shift at the output face with respect to the propagation axis, along the same horizontal direction as the applied field, is reported on figure \ref{fig:bending}.
We observe that   beam bending is negligible for sufficiently low and high intensities, that is, when self-focusing is absent. Furthermore, the maximum beam shift is obtained precisely at the intensity at which self-focusing is most efficient.

\SubSection{Dark irradiance evidence}

The curves reported on figure \ref{fig:cloche}  resemble the photorefractive soliton existence curve ---giving the soliton width as a function of the beam peak intensity to dark irradiance ratio--- as predicted earlier~\cite{chr95josab1628,fre96pre6866} in the framework of a single-photo-carrier model. The dark irradiance plays a central role in this theory since it is the ratio of the beam peak intensity to the dark irradiance which determines the steady state soliton width. The dark irradiance is the equivalent uniform optical irradiance which should be shone on the crystal to generate as much photocarriers as thermal excitation does. In other words, shining the dark irradiance on a given sample doubles its conductivity.

If we assume that this one-photo-carrier model ---which will be named \modelone in the following--- describes our experiments in \InP, we can use its predictions to deduce the dark irradiance in \InP at both the wavelengths we have used. According to it, the minimum width soliton which corresponds to the maximum self-focusing effect is obtained for a beam peak intensity equal to 3 times the dark irradiance, as confirmed recently~\cite{del06josab2323} in a (1+1)D model. Using this hypothesis, the dark irradiance in our sample can then be evaluated, from figure \ref{fig:cloche}, to approximately $70\Milli\Watt/\cm^2$ for the \NIR wavelength and  $2.5\Watt/\cm^2$ for \FIR, at a temperature of $20\Celsius$; the accuracy of the value measured at \NIR is low because of \InP strong absorption at this wavelength. Let us note here that, in both cases, the estimated dark irradiance and the resonance intensity are on the same order of magnitude.

At this stage, one may wonder why \modelone should describe photorefractivity in \InP since both holes and electrons are known to play a significant role~\cite{pic89ol1362,Pic89ap3798} in this semiconductor ---holes are photo-excited and electrons are thermally generated--- whereas \modelone assumes only one carrier, both photo and thermally excited. One may hastily think that electrons can simply be neglected in our experiments. This is however not the case since both the dark irradiance and the resonance intensity are included in our measurement range: they respectively correspond to the  equality of the concentrations on the one hand and of the generation rates on the the other hand, of holes and electrons.

Therefore, since \modelone assumptions do not match with our experiments, we now need to ascertain by an independent measurement that the values of the dark irradiance we have deduced from figures \ref{fig:cloche} and \ref{fig:bending} are indeed correct. We have thus shone a uniform intensity on our \InP sample and measured its photo-conductivity. This allowed us to deduce a $50\Milli\Watt/\cm^2$ dark irradiance at \NIR and a $2\Watt/\cm^2$ one at \FIR, confirming the previous evaluation at this wavelength. The slight discrepancy is accounted for by the relative precision with which figure \ref{fig:cloche} allows the determination of the dark irradiance on the one hand and on the other hand by the difficulty to shine a strong and uniform light on the whole sample for the independent measurement.% Independent measurements at \NIR have not been made owing to apparatus restrictions but ongoing physical processes are the same at both wavelengths  and the conclusion should not differ.

These recent experimental results evidence the fact that the dark irradiance plays a crucial role in photorefractive self-focusing in \InP, as it does in previously developed models, despite the greater complexity of the photorefractive mechanisms in \InP. However, as previously explained, \modelone does not allow to explain our observations, nor do previously published ones~\cite{Cha96ol1333,cha97apl2499,uzd01ol1547}, as will be detailed later on. Therefore, we will now tackle the theoretical problem of showing how and why the dark irradiance plays a significant role in \InP.

\section{Space charge field and dark irradiance}
%\SubSection{Two carrier Kukhtarev model}
Let us start with  the general two-carrier band transport model~\cite{Idr87oc317,uzd01ol1547} reduced to one dimension $x$, which we will consider at steady state in the following. It is described by the set of equations below:
\begin{subequations}
\label{eq:Kukhtarev}
\begin{align}
\frac{\partial E}{\partial x}& =\frac{e}{\epsilon}(N_{D}-N_{A}+p-n-n_{T}),\label{eq:K:Poisson}\\
 j_{n}&=e\mu_{n}nE+\mu_{n}k_{B}T\frac{\partial n}{\partial x},\label{eq:K:jn}\\
 j_{p}&=e\mu_{p}pE-\mu_{p}k_{B}T\frac{\partial p}{\partial x},\label{eq:K:jp}\\
 \frac{\partial n}{\partial t}&=e_{n}n_{T}-c_{n}np_{T}+\frac{1}{e}\frac{\partial j_{n}}{\partial x},%=0
\label{eq:K:n}\\
  \frac{\partial p}{\partial t}& =e_{p}p_{T}-c_{p}pn_{T}-\frac{1}{e}\frac{\partial j_{p}}{\partial x},%=0
\label{eq:K:p}\\
 \frac{\partial n_{T}}{\partial t}&=e_{p}p_{T}-e_{n}n_{T}-c_{p}pn_{T}+c_{n}np_{T},%=0
\label{eq:K:nT}\\
N_{T}& =n_{T}+ p_{T},
\end{align}
\end{subequations}
where $E$ is the electric field, $n$ and $p$ are the electron and
hole densities in the respective conduction and valence bands,
$n_{T}=\Fe 2$ is the density of ionized occupied traps,
$p_{T}=\Fe 3$ is the density of neutral unoccupied traps,
$j_{n}$ and $j_{p}$ are  the electron and hole
current densities, respectively. $N_{T}$, $N_{D}$ and $N_{A}$ are   the
densities of iron atoms, the shallow donors and the shallow
acceptors, respectively. The charge mobilities are given by $\mu_{n}$ for
electrons and $\mu_{p}$ for holes, the electron and hole
recombination rate are respectively  $c_{n}$ and $c_{p}$ , $T$ is
the temperature and $k_{b}$ is the Boltzmann constant. The
dielectric permittivity is given by $\epsilon$ while $e$ is the
elementary charge.% $V_{app}$ is the voltage  applied externally to the  crystal of width $d$.
The electron and hole
generation rates  $e_{n}$ and $e_{p}$  depend on both
thermal and optical emission as
described by% equations (\ref{elec}) and (\ref{trou}) :
\begin{subequations}
\label{eq:rates}
	\begin{align}
	e_{n}&=e^{th}_{n}+\sigma_{n}I\approx e^{th}_{n},\label{eq:rate:elec}\\
	e_{p}&=e^{th}_{p}+\sigma_{p}I\approx\sigma_{p}I\label{zq:rate:trou}
	\end{align}
\end{subequations}
where the thermal contribution to the emission rate coefficient is
$e^{th}$ and the optical cross section of the carriers is given by
$\sigma$, $I$ is the spatially dependent intensity of light. The approximations in both above equations are valid in the case of \InP as stated above, with thermally excited electrons and photo-excited holes.
%Further approximations will be considered.
 The need for a large applied electric field leads us to neglect carrier diffusion when compared to drift, diffusion which is accounted for by all the terms which contain $k_BT$.

Solving the differential system (\ref{eq:Kukhtarev}) requires defining the boundary conditions. To that aim, we will consider that the input beam is small with respect to the sample dimensions and thus that its influence is limited to its surroundings. Therefore, the boundary values for $E$, $n$, $p$, $n_T$, $p_T$, $j_n$ and $j_p$ are the values reached at steady state in the dark.
As was previously done~\cite{Idr87oc317,pic89ol1362,Pic89ap3798,khel06oc169}, they can be deduced from 
\TextColor{blue}{system (\ref{eq:Kukhtarev}), % neglecting
setting to zero} all derivatives and optical intensity. Following, if the $0$ subscript identifies the boundary condition: $\Bound E=\Frac{V}{d}$ where $V$ is the applied voltage and $d$ the crystal width; $\Bound{n}=\Dark{n}{p}$ and $\Bound{p}=\Dark{p}{n}$ where ${n}_{T0}=N_D-N_A$ and $p_{T0}=N_T-{n}_{T0}$ are the ionised and non-ionised iron densities in the dark~\cite{pic89ol1362,Pic89ap3798}; $j_n=\Current{n}$ and $j_p=e\mu_ppE_0$.

\TextColor{blue}{
The time evolution of the electric field $E$ is determined as follows. The time derivative of
 Eq.  (\ref{eq:K:Poisson})  is simplified using Eqs.  (\ref{eq:K:n},\ref{eq:K:p},\ref{eq:K:nT}) to yield
\begin{equation}
\frac{\partial^2E}{\partial t\partial x}=
\frac{-1}\varepsilon\frac{\partial\left(j_n+j_p\right)}{\partial x}.
\end{equation}
Making use of Eqs.  (\ref{eq:K:jn},\ref{eq:K:jp}) and integrating with respect to the space variable $x$, it becomes
\begin{equation}
\frac{\partial E}{\partial t}=\frac{-e}\varepsilon\left[\left(\mu_nn+\mu_pp\right)E.
-\left(\mu_{n}n_0+\mu_{p}p_0\right)E_0\right]\label{dte}
\end{equation}}

%With these approximation,
\TextColor{blue}{At steady state, the evolution equations (\ref{eq:K:n}), (\ref{eq:K:p}), (\ref{eq:K:nT}) and (\ref{dte}) reduce to}
\begin{subequations}\label{eq:K:simp}
\begin{align}
\left(\mu_{n}n+\mu_{p}p\right)E-\left(\mu_{n}n_{0}+\mu_{p}p_{0}\right)E_{0}&=0,\label{eq:K:simp:E}\\
e_{n}n_{T}-c_{n}n\left(N_{T}-n_{T}\right)+\mu_{n}n\frac{\partial E}{\partial x}+\mu_{n}E\frac{\partial n}{\partial x}&=0,\label{eq:K:simp:n}\\
e_{p}\left(N_{T}-n_{T}\right)-c_{p}pn_{T}- \mu_{p}p\frac{\partial E}{\partial x}-\mu_{p}E\frac{\partial p}{\partial x}&=0.\label{eq:K:simp:p}%\\
%n_{T_{0}}-\frac{\epsilon}{e}\frac{\partial E}{\partial x}&=n_{T}\label{eq:K:simp:nT}
\end{align}
\end{subequations}

To further simplify this differential system, the light intensity spatial variations will be assumed small enough so that the resulting free carriers spatial variations
\TextColor{blue}{remain small, and the derivatives  ${\partial n}/{\partial x}$ and  ${\partial p}/{\partial x}$ can be neglected in
 equations (\ref{eq:K:simp:n}) and (\ref{eq:K:simp:p}). 
In the same spirit, we also assume that the field derivative ${\partial E/\partial x}$  in the same equations can be neglected,
 precisely $\partial E/\partial x\ll e_n^{th}N_T/\mu_n n_0$.}

\TextColor{blue}{The left-hand-side of Eq.  (\ref{eq:K:Poisson}) becomes zero within this approximation.}
\TextColor{green}{ Light intensity and temperature will again be considered not so large so that free carriers density remain small with respect to ionised iron density: $n\ll n_T$ and $p\ll n_T$, so that equation (\ref{eq:K:Poisson}) } %can be re-written as
%\begin{equation}
% \frac{\partial E}{\partial x} =\frac{e}{\epsilon}(N_{D}-N_{A}-n_{T})
%\end{equation}
\TextColor{blue}{reduces to $n_T=n_{T0}$, and consequently from Eqs. (\ref{eq:K:simp:n}) and  (\ref{eq:K:simp:p})  we get $n=n_0$ and $p=\left(\sigma_p I p_{T0}\right)/\left(c_p
n_{T0}\right)$.
The assumptions made are valid for intensities small against ${c_pn_{T0}^2}/{\sigma_pp_{T0}}$, itself greater than the $\mathrm{TW}/\mathrm{cm^2}$, and at temperatures orders of magnitude above room temperature.
Within these approximations,  we obtain} the electric field $E$ dependency upon the local intensity $I$:
\begin{equation}
 \label{EvsI}
	E=\frac{\Idark}{\Idark + I} E_0.
\end{equation}

This expression is strictly the same as the one found before in the framework of \modelone~\cite{chr95josab1628,fre96pre6866} when diffusion is neglected; only the expression of the dark irradiance $\Idark$ differs (the approximation below being valid in the case of \InP):
\begin{equation}
 \Idark=\frac{e^{th}_pp^2_{T0}c_n\mu_p+e^{th}_nn^2_{T0}c_p\mu_n}{\sigma^0_pp^2_{T0}c_n\mu_p+\sigma^0_nn^2_{T0}c_p\mu_n}
\approx\frac{e^{th}_nn^2_{T0}c_p\mu_n}{\sigma^0_pp^2_{T0}c_n\mu_p}.
\end{equation}

Let us point out here that this has implications beyond the above equation: all the conclusions that have been driven, in \modelone, from an equation identical to (\ref{EvsI}), still hold, if the above value of $\Idark$ is accounted for. There is for instance no need to derive again the soliton existence curve since the involved equations are the same. Therefore and using the predictions \modelone, our observations of self-focusing in \InP are partially explained. Indeed the presence of self-focusing over a broad range of intensity is predicted and no electron-hole resonance effect  is observed. The beam bending is however not yet accounted for, which is one reason to further extend and refine our model.
% but this was  expected since we neglected diffusion at the start.

\section{Resonance intensity}

As was said before, we felt the need to establish the theoretical analysis presented in the previous section ---which we will now call \modeltwo--- because we could not satisfactorily explain our observations using the theory available in the literature~\cite{Cha96ol1333,uzd01ol1547} for one particular reason: this theory ---which we will call \modelthree--- predicts the existence of a resonance intensity around which the self-focusing phenomenon reverses itself; while we have successfully measured the resonance intensity in our samples, we were never able to measure the predicted reversal from focusing to defocusing.%, despite a wide range of intensities used, despite several dopings of several samples from several origins.

The resonance intensity around which a self-focusing behavior reversal is predicted by \modelthree is precisely the same resonance intensity at which TWM gain is maximum.  As was said before, the resonance intensity is also the intensity at which the generation rates of holes and electrons are equal:
%The origin of the behavior reversal predicted by \modelthree can be interpreted by an analogy with the Two Wave Mixing process and particularly with the amplitude and phase of the space charge field developed in \InP illuminated by  a sinusoidal pattern of light, issued for instance from a two beam interference. In that case, as we have measured, there is a particular intensity at which the generation rates of electrons and holes are equal,
% which is precisely the resonance intensity:

\begin{equation}
 \Ires=\frac{e^{th}_p p_{\Bound{T}}+e^{th}_n n_{\Bound{T}}}{\sigma^0_p p_{\Bound{T}}+\sigma^0_n n_{\Bound{T}}}\approx\frac{e^{th}_n n_{\Bound{T}}}{\sigma^0_p p_{\Bound{T}}}.
\label{ires}
\end{equation}

At resonance, the imaginary part of the space charge field reaches a maximum while the real part ---the \emph{in phase} component--- reaches zero. Following, the real part changes sign around $\Ires$~\cite{pic89ol1362,Cha96ol1333}. The link with self-focusing can then be thought of as follows: the in phase component of the space-charge field ---the real part--- is a \emph{local} photorefractive response, responsible for self-focusing or defocusing, whereas the out-of phase one --the imaginary part--- is responsible for beam bending.  This phenomenological interpretation has been confirmed by an experiment~\cite{Cha96ol1333}. Later on,  analytical derivations and numerical simulations~\cite{uzd01ol1547} came to the same predictions. As stated before, however, we were never able to reproduce the experiment, for a reason we will now try to give hints of.

First of all, in the case of \InP, there is a close and fairly simple relationship between $\Idark$ and $\Ires$:

\begin{equation}
\frac{\Idark}{\Ires}=\frac{\mu_nc_p}{\mu_pc_n}\times\frac{n_{T0}}{p_{T0}}.
\label{eq:ratio}
\end{equation}

We have evaluated the above ratio to $0.65$ with values issued from the literature~\cite{AFR74ssc59,suz84jap291,Ozksab1997}, a numerical value which is fully compatible with those we have measured at \FIR, the discrepancy for the measures at \NIR being accounted for by the strong absorption impairing the $\Idark$ estimation from figures \ref{fig:cloche} and \ref{fig:bending}.

The above ratio (\ref{eq:ratio}) evidences the ratio $\Frac{\Fe 2}{\Fe 3}$ in the dark (${n_{T0}}/{p_{T0}}$). Therefore, the oxydo-reduction state plays a significant role, as well as the doping of the samples, in the ratio of the two characteristic intensities. Therefore, a different doping but also a different $\Frac{\Fe 2}{\Fe 3}$ ratio might be the cause of the discrepancy between our experimental results, which do not exhibit a crucial role for the resonance intensity, and those previously reported.

However, our theoretical analysis, \modeltwo, does not either bring the resonance intensity forth. A more universal model is in development in order to predict a doping dependent behavior and thus unify \modeltwo and \modelthree.
%We believe that the reason lies in the fact that, in TWM theory and experiments, the resonance intensity is reached when the \emph{average} of the input light is equal to $\Ires$. On the contrary, in a single beam experiment as we have done, the average intensity over the whole sample is generally close to $0$ and never reaches $\Ires$, keeping always on the same side of the resonance, and thus inducing no behavior reversal.

\section{Conclusion}
In the light of recent experimental results, we have shown that despite more complex photorefractive charge transport mechanisms, steady-state photorefractive self-focusing and solitons in \InP can be described by the same model as was done for the simpler one-free-carrier model, provided diffusion is neglected. We have evidenced in the process the role of the dark irradiance. Our model does not, however, evidence any crucial role for the resonance intensity, a role which is to be the subject of future studies.

\section{Acknowledgements}
The authors wish to thank the Lorraine R\'egion for their financial support.
\bibliographystyle{apsrev}
\bibliography{INP,PRsolitons,PubliERPO}

\end{document}